# Giant exciton binding energy in bulk $CrCl_3$

*Georgy Ermolaev[1†], Tagir Mazitov[2†], Anton Minnekhanov[1], Arslan Mazitov[1], Gleb Tselikov[1], Aleksandr Slavich[1], Mikhail Tatmyshevskiy[2], Mikhail Kashchenko[2], Nikolay Pak[1], Andrey Vyshnevyy[1], Valentin Solovei[1], Ivan Kazantsev[1], Gleb Tikhonowski[1], Alexander Melentev[2], Elena Zhukova[2], Dmitriy Grudinin[1], Junhua Luo[3], Ivan Kruglov[1], Aleksey Arsenin[1], Sangen Zhao[3,4*], Kostya S. Novoselov[5,6,7*], Andrey Katanin[2*], Valentyn S. Volkov[1*]*

[1]*Emerging Technologies Research Center, XPANCEO, Internet City, Emmay Tower, Dubai, United Arab Emirates*

[2]*Moscow Center for Advanced Studies, Kulakova str. 20, Moscow, 123592, Russia*

[3]*State Key Laboratory of Structural Chemistry, Fujian Institute of Research on the Structure of Matter, Chinese Academy of Sciences, Fuzhou, Fujian, 350002 P. R. China*

[4]*Quantum Science Center of Guangdong–Hong Kong–Macao Greater Bay Area (Guangdong), Shenzhen 518045, P. R. China*

[5]*National Graphene Institute (NGI), University of Manchester, Manchester, M13 9PL, UK*

[6]*Department of Materials Science and Engineering, National University of Singapore, Singapore, 03-09 EA, Singapore*

[7]*Institute for Functional Intelligent Materials, National University of Singapore, 117544, Singapore, Singapore*

[†]These authors contributed equally to this work

**Correspondence should be addressed to e-mail:* zhaosangen@quantumsc.cn, kostya@nus.edu.sg, andrey.katanin@gmail.com, and vsv@xpanceo.com

## Abstract

**Van der Waals (vdW) materials, with their unique combination of electronic, optical, and magnetic properties, are emerging as promising platforms for exploring excitonic phenomena. Thus far, the choice of materials with exceptional excitonic response has been limited to two-dimensional (2D) configurations of vdW materials. At the same time, large interlayer distance and the possibility to create a variety of heterostructures offers an opportunity to control the dielectric screening in van der Waals heterostructures and van der Waals 3D materials, thus engineering the excitonic properties. Here, we reveal that bulk vdW crystal $CrCl_3$ answers this quest with a record exciton binding energy of 1.64 eV owing to a delicate interplay of quasi-2D electronic confinement and local magnetic correlations. We also suggest that the non-local magnetic correlations play an important role in the temperature dependence of photoluminescence intensity. Furthermore, we observe colossal binding energies in vdW crystals $NbOCl_2$ (0.66 eV) and $MoCl_3$ (0.35 eV) and formulate a universal exciton binding energy dependence on bandgap for 2D and 3D vdW materials. Hence, our findings establish a fundamental link between the layered structure of vdW materials and their excitonic properties.**

## Introduction

Excitons, the bound state of electrons and holes, have led to myriad phenomena and applications, particularly with the emergence of two-dimensional (2D) materials[1,2]. These materials demonstrate giant exciton binding energy due to their 2D nature, which allows their efficient use in photodetectors[3], light-emitting diodes[4], photovoltaics[5], quantum communication and computing[6,7], valleytronics[8], and exciton-polariton transport[9]. In light of this, an intriguing question emerges: if layered materials, essentially stacks of monolayers, also exhibit these fundamental excitonic features, can we anticipate equally enormous binding energies in bulk layered materials? This hypothesis suggests that the unique excitonic behavior observed in 2D materials might extend to their bulk counterparts, potentially opening new avenues for optoelectronics and quantum technologies[6,10]. Numerous groups[11–13] tried to answer this question by investigating bulk transition metal dichalcogenides, but their findings are controversial. This controversy stems from the experimental difficulties arising from the overlapping of excitonic peaks with quasiparticle bandgap[14,15], which makes the task of measuring binding energy a challenging problem. Hence, one needs clearly separated excitons from bandgap to determine their binding energy unequivocally. In this regard, the new family of magnetic halogen-based layered materials, including $CrCl_3$, $CrBr_3$, and $CrI_3$, is a promising platform to resolve this challenge since, for them, theory predicts individual excitonic peaks even in the bulk form[16,17]. Moreover, their pronounced magnetic properties[18–20] may positively affect their excitonic response. Therefore, a van der Waals (vdW) crystal with giant exciton binding energy originating from magnetic order is highly desirable for developing exciton-based devices.

In this work, we experimentally and theoretically observed giant exciton binding energy in bulk $CrCl_3$. Our results demonstrate that excitons in $CrCl_3$ originate from local magnetic correlations even at room temperature despite the small Néel temperature of about 20 K[19–21], while their temperature dependence is affected by the short range magnetic order. We also emphasize that giant exciton binding energy in this material is due to the quasi-two-dimensionality of its electronic properties. Furthermore, we found other vdW crystals, $NbOCl_2$ and $MoCl_3$, with similarly large exciton binding energy, which is also related to their layered structure. As a result, our work exposes a non-trivial link between electronic, magnetic, and optical properties and shows that halogen-based vdW materials are a promising material platform for comprehensive exciton physics and applications.

## Results

### Excitons in different dimensionalities

In 2D materials (Figure 1a), excitons exhibit significantly enhanced binding energies compared to their three-dimensional (3D) counterparts (Figure 1b)[14,15]. This phenomenon can be attributed to several key factors intrinsic to the reduced dimensionality. Firstly, the reduced dielectric screening in 2D materials plays a crucial role. In 3D material, the dielectric permittivity is relatively high, leading to effective screening of the Coulomb interaction between the electron and hole that form the exciton. However, in the 2D case, the dielectric environment has a relatively low dielectric screening, facilitating the Coulomb interaction (Figure 1a). Secondly, the quantum confinement effect in 2D materials enhances excitonic binding energy. Additionally, in 2D, Coulomb potential modifies from the Gauss $1/r$ law to the Rytova–Keldysh form[15], further strengthening electron-hole binding. In this context, the most interesting case for excitons is layered/vdW materials (Figure 1c), where excitons are also confined within a single plane, sharing similarities with 2D material (Figure 1a). At the same time, layered/vdW materials have a relatively strong dielectric response, making it look like a 3D case (Figure 1b). Consequently, excitons in layered materials should share properties from both 2D and 3D situations with an intermediate value of binding energies. Naively, one could expect that vdW crystals with a relatively large dielectric response ($\varepsilon \sim 15$) would exhibit a moderate binding energy, while for the relatively small dielectric response ($\varepsilon \sim 5$), one can anticipate a colossal binding energy in vdW materials comparable to 2D materials.

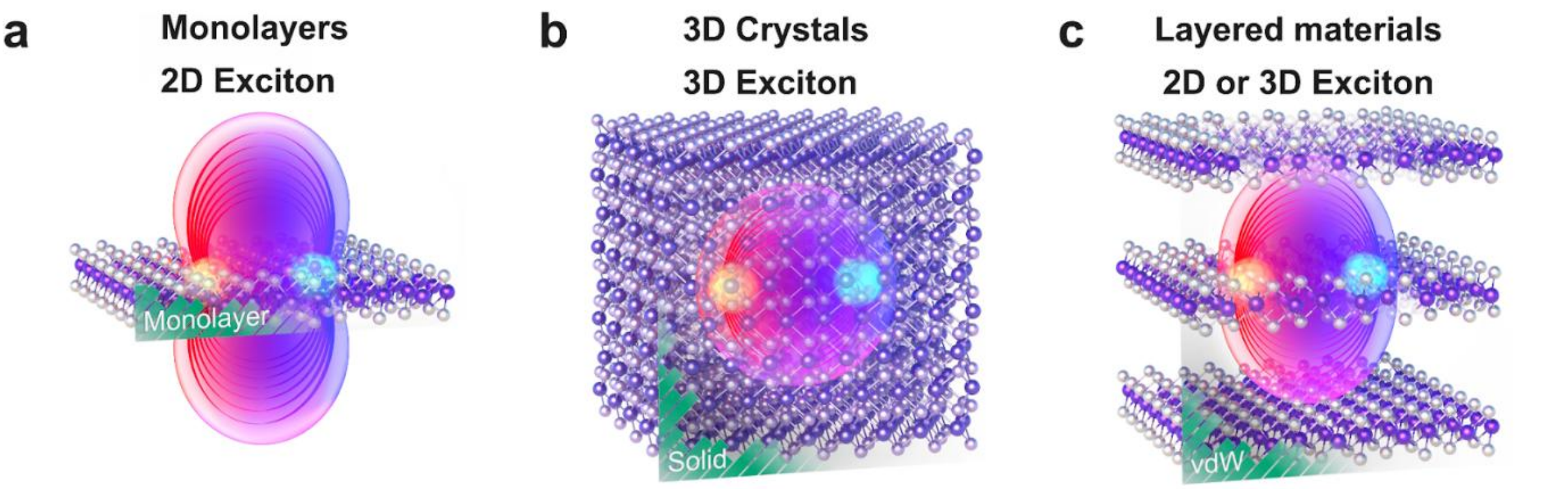

**Figure 1. Real-space representations of excitons in different dimensions. a,** Exciton in 2D material. **b,** Exciton in 3D material. **c,** Exciton in layered bulk materials, where it can demonstrate both 2D and 3D features.

### Observation of $CrCl_3$ excitons

To validate the proposed concept, we first focus on $CrCl_3$ excitons, which recently received considerable attention[16,17,19,22]. In layered form, $CrCl_3$ is a vdW monoclinic crystal with the following lattice parameters: $a$ = 0.59588 nm, $b$ = 1.032206 nm, $c$ = 0.61138 nm, $\alpha$ = 90°, $\beta$ = 108.495°, and $\gamma$ = 90° (Figure 2a) in the C2/m phase observed at room temperature[23]. Its reduced crystal symmetry generally leads to a dielectric tensor of non-diagonal form with wandering principal optical axes. Fortunately, its in-plane optical anisotropy is negligible (Supplementary Note 1), allowing us to treat $CrCl_3$ as an optically uniaxial system with dielectric tensor matrix diag($n_{ab}+ik_{ab}$, $n_{ab}+ik_{ab}$, $n_c+ik_c$), where $n_{ab}$, $n_c$ and $k_{ab}$, $k_c$ are refractive indices and extinction coefficients along *ab*-plane and *c*-axis, respectively.

To obtain anisotropic optical constants of $CrCl_3$, we measured reflectance (Figure 2b), transmittance (Figure 2c), and ellipsometry (Supplementary Note 2) spectra of exfoliated $CrCl_3$ flake on the glass substrate. Notably, the resulting spectra demonstrate pronounced Fabry-Perot resonances with apparent dips around 1.7 eV and 2.3 eV in transmittance, indicating the presence of excitons (Figure 2c). For simultaneous fitting of reflectance, transmittance, and ellipsometry spectra, we leverage standard for layered materials Tauc–Lorentz oscillators approach for in-plane optical constants and a Cauchy model for out-of-plane optical constants of $CrCl_3$ (Supplementary Note 3)[9]. The resulting refractive indices and extinction coefficients are displayed in Figures 2d-e. In accordance with theoretical studies[17], which predict A- and B-excitons positions at 1.78 eV and 2.51 eV, respectively, we observe them at $E_A$ = 1.67 eV and $E_B$ = 2.28 eV (the inset in Figure 2e). More importantly, the bandgap of $CrCl_3$ is well separated from excitonic peaks, enabling unambiguous determination of bandgap position $E_g$ = 3.31 eV (Figure 2e) and the corresponding excitons binding energies $E_A^{\mathrm{binding}} = E_g - E_A = 1.64$ eV and $E_B^{\mathrm{binding}} = E_g - E_B = 1.03$ eV. To the best of our knowledge, these values are the largest among all experimentally measured excitonic materials, even compared to previous record-holders monolayers[14,15] $MoS_2$ and $WS_2$ with $E_A^{\mathrm{binding}} = 0.45$ eV and $E_B^{\mathrm{binding}} = 0.32$ eV, respectively.

Apart from intriguing linear optical properties, excitons in $CrCl_3$ demonstrate distinctive photoluminescence (PL), as seen in Figures 2f-h. Unlike the isotropy of in-plane optical constants (Supplementary Note 1), PL exhibits an anisotropic response (Figures 2e-f), allowing us to determine the preferential orientation of the fundamental exciton with respect to the crystallographic axes. It turns out that the fundamental exciton in $CrCl_3$ is rotated by 48° with respect to the crystallographic *a*-axis (Supplementary Note 4). Furthermore, at both standard excitation wavelengths, 532 nm (Figure 2f) and 785 nm (Figure 2g), $CrCl_3$ displays a similar PL pattern (Figure 2h), even though they correspond to different excitons: the corresponding laser excitation energies 2.33 eV and 1.58 eV coincide well with the B- (2.28 eV) and A-exciton (1.67 eV) positions, respectively. This behavior suggests that the B-exciton relaxes to the A-exciton before the radiative recombination. In addition to PL, $CrCl_3$ demonstrates promising waveguiding properties, with a propagation length of 70 µm for exciton-polaritons, which is an order of magnitude larger than that of traditional $MoSe_2$ exciton, polaritons with a propagation length of 2.5 µm (Supplementary Note 5).

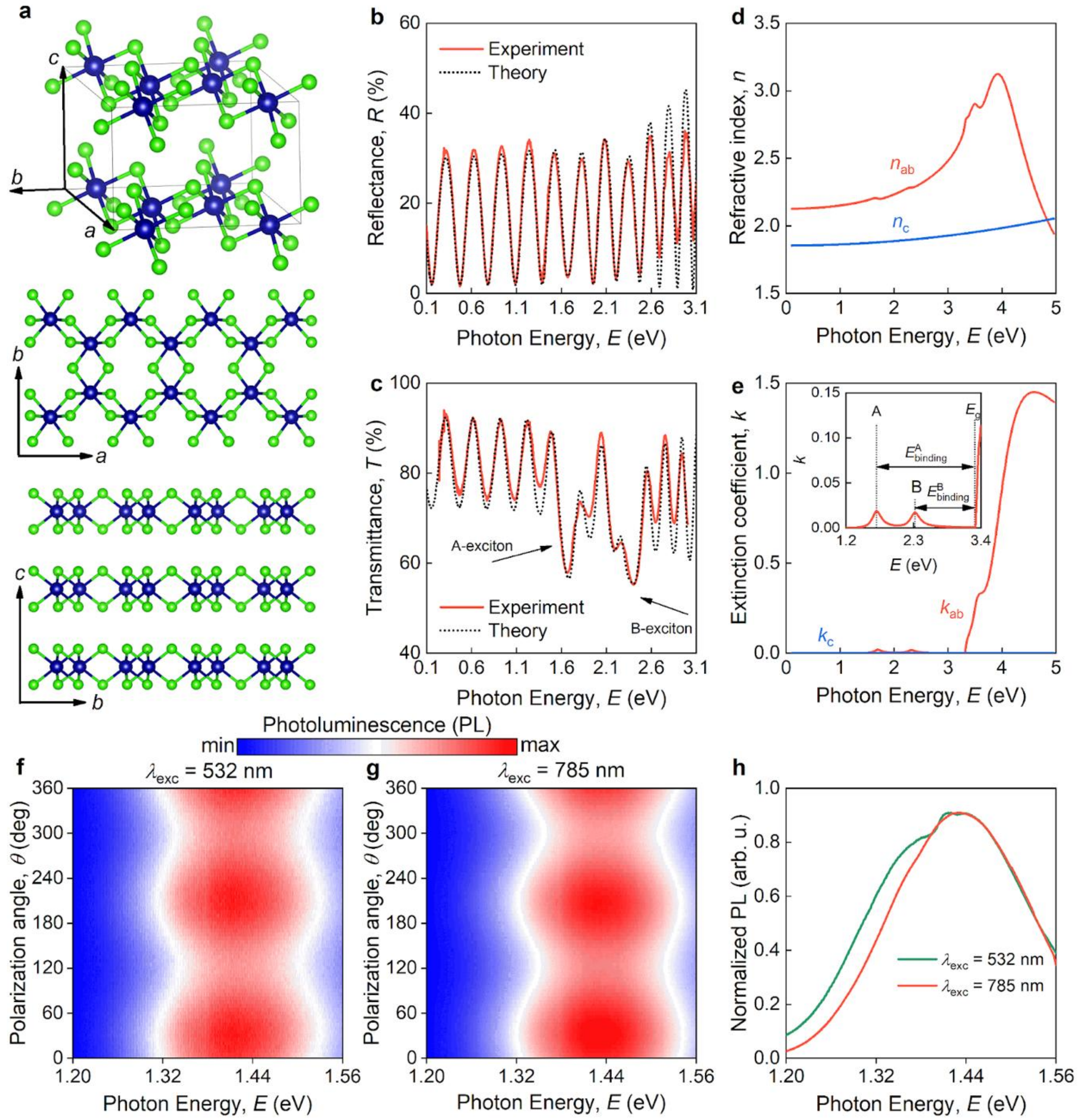

**Figure 2. Excitonic optical properties of $CrCl_3$. a,** Crystal structure of $CrCl_3$: three-dimensional view of the unit cell (top), view along the *c*-axis (middle), view along the *a*-axis (bottom). **b,** Reflectance and **c,** transmittance spectra of 945-nm thick $CrCl_3$ flake. **d,** Anisotropic refractive index of $CrCl_3$. **e,** Anisotropic extinction coefficient of $CrCl_3$. The inset shows the spectral region with excitons. Photoluminescence spectra of $CrCl_3$ for **f,** 532 nm and **g,** 785 nm excitation wavelengths. **h,** Averaged over polarization angles normalized photoluminescence spectra of $CrCl_3$.

## The effects of magnetic correlations on $CrCl_3$ excitons

In order to further analyze the optical properties of $CrCl_3$, especially its excitonic properties and dielectric response in the presence of magnetic correlations, we calculated the optical constants within the $G_0W_0$ approximation with spin-orbit coupling, and further accounted for the two-particle (electron-hole) interactions using a Bethe-Salpeter equation (BSE) formalism. Following experimental data[18,19] and other recent studies[16,20] of the $CrCl_3$ system, we consider the out-of-plane ferromagnetic (FM-OOP) order of the Cr spins to be representative of the actual spin configuration at room temperature. In the long-wavelength limit, our $G_0W_0$ results on refractive index and extinction coefficient (Figure 3a) demonstrate very good quantitative agreement with the experimental results obtained with ellipsometry. However, the lack of two-particle interactions in the $G_0W_0$ approach limits the expressivity in the short-wavelength region. Indeed, this picture changes quite noticeably when taking the electron-hole interaction into account within BSE formalism: our results on the position of the A exciton ($E_A$ = 1.70 eV) are aligned perfectly with the experimental results obtained in this work (Figure 3b). The B exciton is less pronounced in BSE and can be noticed by the inflection of the absorbance curve around $E_B$ = 2.40 eV, which is slightly higher than the experimental result $E_B$ = 2.28 eV (Figure 3b). Additionally, both results agree on a quantitative level with earlier experimental[24,25] and theoretical[16,17] works (Figure 3c).

Notably, even in the presence of spin-orbit interaction, not only the considered FM-OOP, but also in-plane ferromagnetic (FM-IP) and in-plane antiferromagnetic (AFM-IP) configurations yield an excitonic peak at ~1.7 eV, as we demonstrate in Supplementary Note 6. Indeed, all three magnetic configurations demonstrate a similar flat and symmetric electronic dispersion with comparable band gaps (see Supplementary Note 6). This observation aligns with previous work[17], which considered the AFM-IP configuration of the Cr spins, yet the position of the A exciton agrees with our results.

As mentioned above, $CrCl_3$ is not magnetically ordered at room temperature[20] and becomes antiferromagnetic only at temperatures of about 20 K. However, local magnetic correlations persist already at room temperature and have profound effects on opening the gap in electronic spectrum and appearance of excitons even in the absence of long-range magnetic order. In order to estimate the strength of both local and non-local magnetic correlations, we evaluated the local and non-local static susceptibility, as well as the corresponding exchange interactions, within the DFT+DMFT approach[26]. The evidence of the presence and substantial impact of local magnetic correlations on the electronic and optical properties of the material is obtained from the distribution of the spectral density of electronic states in Figure 3d, as well as the temperature evolution of the local spin susceptibility (see Supplemental Note 7). In particular, strong correlations, arising in part from the Hund's exchange of the Cr d-states, play a decisive role in the presence of the gap in the electronic spectrum and the robustness of the magneto-optical properties against temperature variations. Remarkably, the spectral density of the d-orbitals of Cr atoms in $CrCl_3$ from our DFT+DMFT calculations align closely with the band structure obtained within the $G_0W_0$ approach. In particular, Figure 3d demonstrates that the spectral density reaches its maximum at the same bands that contribute most to the formation of the A exciton. These results allow us to assert that local spin correlations (governed by Hund exchange) constitute the origin of the gap in the electronic spectrum, and, therefore, the origin of the excitons in the considering material.

Additionally, we analyzed the contributions from the various neighbors in the non-local susceptibility and exchange interactions (see Supplementary Note 7 for more details). From Figure 3e one can see that the exchange interactions (which are determined by the non-local part of magnetic susceptibility) are largely limited to the nearest neighbors, with substantial attenuation already observed at the second neighbors. This also correlates with the spatial distribution of the exciton wave functions (see Figure 3f), since both quantities are determined by electron and hole propagation in the considered substance. At the same time, this clearly demonstrates intertwining of magnetic and optical properties of $CrCl_3$. The obtained exchange interactions at the nearest neighbor sites, as well as the corresponding contributions to the non-local susceptibility (Figure 3g) are all positive, which indicates ferromagnetic short-range order, in agreement with the onset of ferromagnetic long-range order with lowering the temperature. The key feature is a rapid increase of the non-local susceptibility (Figure 3g) with decreasing temperature. This result also agrees well with the strong increase of the linewidth of electron spin resonance with decrease of temperature in $CrCl_3$.[27] At room temperature, the value of the in-plane nearest-neighbor contribution (per neighbor) is already $\chi_{nn}$ ~ 12 $eV^{-1}$. Although the vertex and non-local self-energy corrections can suppress this contribution, it is expected to affect the optical properties, making the electronic excitations on nearest neighbor Cr sites more coherent. The correlations of the nearest Cr-atoms in neighboring layers are much smaller than between atoms within the same layer, suggesting the quasi-two-dimensionality of this material.

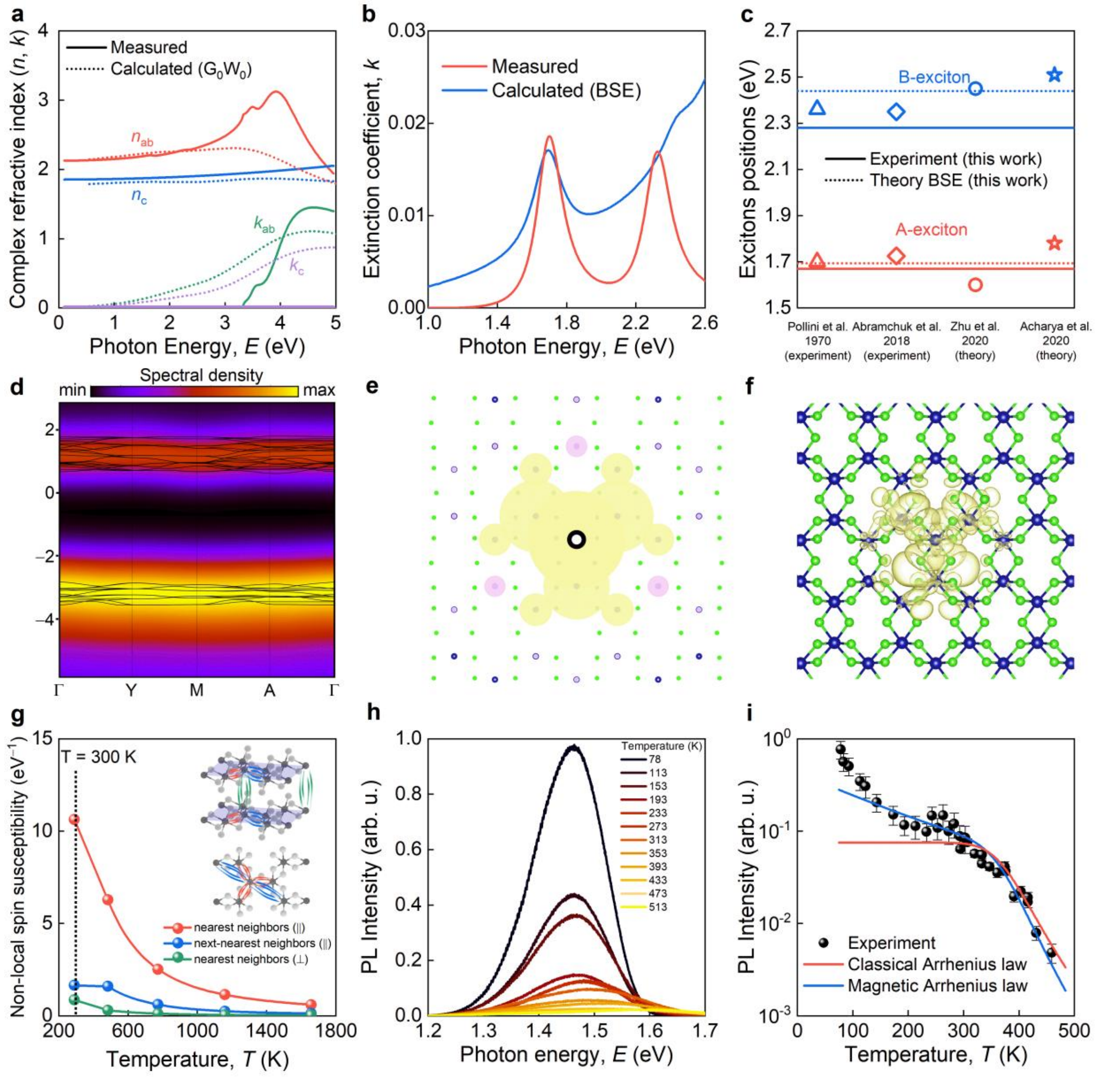

**Figure 3. Theory of magnetic excitons in $CrCl_3$.** **a**, Real and imaginary parts of the complex refractive index, measured experimentally (solid lines) amd calculated within the $G_0W_0$ approach (dotted lines). **b,** Comparison of the exciton peaks in the extinction coefficient, measured experimentally (red line) and calculated within the Bethe-Salpeter Equation (BSE) framework (blue line). **c,** Comparison of A- and B-exciton positions obtained in this work with existing works on the optical response of $CrCl_3$. **d,** Band structure in GW approach for the out-of-plane (FM) (black lines), compared to the spectral density of states in DFT+DMFT approach for $CrCl_3$ (color scale). **e,** exchange interactions (ferromagnetic interaction is shown by yellow circles, antiferromagnetic by purple circles; sizes of circles demonstrate the value of interaction with the central atom). **f,** Real space distributions in ab-plane of wavefunctions of electrons and holes in exciton. **g,** Contributions (per neighbor) to the non-local spin susceptibility from the nearest (red line, circles), next-nearest (blue line, circles) neighbors within the plane, and nearest neighbors between the planes (green line with circles). The inset shows the schematics of in-plane ( ll) nearest neighbors and next-nearest neighbors and out-of-plane (⊥) nearest neighbors. **h,** Temperature dependance of photoluminescence spectra of $CrCl_3$. **i,** Temperature dependence of photoluminescence intensity.

To confirm the important role of non-local magnetic correlations on nearest neighbor Cr sites in the properties of excitons in $CrCl_3$, we have also measured the temperature dependence of PL response. Given the rapid change of non-local spin susceptibility with temperature (Figure 3f), we expect a similar dependence of exciton photoluminescence (PL). Indeed, PL intensity significantly increases with the cooling of $CrCl_3$, as seen in Figure 3g. However, even in the classical case, temperature $T$ affects the PL intensity $I$ following Arrhenius law $I(T) = I_0/(1+C\exp[-E_a/(k_B T)])$, where $I_0$ is PL intensity at low temperatures, $C$ is a phenomenological constant, $k_B$ is a Boltzmann constant, and $E_a$ is an activation energy[28,29]. Figure 3h shows that the Arrhenius law describes the PL temperature dependence only for high temperatures, where magnetic correlations are not quite important due to small magnitude of the non-local susceptibility (Figure 3g). Considering Figure 3g and assuming that the low-temperature intensity $I_0$ scales with temperature according to the extrapolated temperature dependence of the obtained non-local spin susceptibility, we arrive at an agreement of the experiment and theory also in the intermediate temperature range (Figure 3h). Therefore, this result confirms the effect of the non-local magnetic correlations on the properties of excitons in $CrCl_3$.

**Excitons in other chalcogen-based vdW crystals beyond $CrCl_3$**

In addition to $CrCl_3$, we expect a similar excitonic response for $CrBr_3$ and $CrI_3$ crystals[16]. However, this behavior is not limited to Cr-based vdW materials. For example, we discovered that $NbOCl_2$ and $MoCl_3$ have giant exciton binding energies, as seen in Figure 4. Indeed, their ellipsometry spectra (Supplementary Note 8 and Supplementary Note 9) demonstrate similar excitonic features as the $CrCl_3$ spectrum in Figure 2c. However, unlike $CrCl_3$, $NbOCl_2$ and $MoCl_3$ exhibit large in-plane optical anisotropy with biaxial dielectric tensors[30]. As a result of ellipsometry fitting, we have obtained anisotropic refractive indices (Figure 4a-b) and extinction coefficients (Figure 4c-d) for $NbOCl_2$ and $MoCl_3$. It is worth noting that a clearly separated from the bandgap excitonic peak for these materials only along one in-plane direction (Figure 4c-d). Based on Figure 4c-d, we determine the positions of excitons ($E_{\text{exciton}}^{\text{NbOCl}_2}$ = 1.84 eV and $E_{\text{exciton}}^{\text{MoCl}_3}$ = 1.55 eV) and corresponding bandgaps ($E_g^{\text{NbOCl}_2}$ = 2.50 eV and $E_g^{\text{MoCl}_3}$ = 1.90 eV). Therefore, binding energies can be found as $E_{\text{binding}}^{\text{NbOCl}_2} = E_{\text{exciton}}^{\text{NbOCl}_2} - E_g^{\text{NbOCl}_2} =$ 0.66 eV and $E_{\text{binding}}^{\text{MoCl}_3} = E_{\text{exciton}}^{\text{MoCl}_3} - E_g^{\text{MoCl}_3} =$ 0.35 eV. Although these values are smaller than for $CrCl_3$ ($E_{\text{binding}}^{\text{CrCl}_3}$ = 1.64 eV), they are still colossal compared to other excitonic materials.

Given these record values, it is instructive to provide a more comprehensive comparison of binding energies since, for example, the bandgap is a crucial parameter that determines exciton binding energy, as seen from the correlation of classical 3D excitonic materials in Figure 5[31]. Obviously, a larger bandgap generally results in larger exciton binding energy (Figure 5). If we place the results for $CrCl_3$, $NbOCl_2$, and $MoCl_3$ alongside monolayers $MoS_2$[14] and $WS_2$[15] on the same graph, we can notice that their binding energies are comparable to each other. Still, they are definitely larger than those for standard 3D excitonic materials. Of immediate interest is bulk $MoS_2$[12], which is in between the classical "3D exciton" region and our "2D exciton" region. This effect most likely originates from the high dielectric permittivity of 3D $MoS_2$ ($\varepsilon \sim 16$), which effectively screens the Coulomb interaction (see the above section "Excitons in different dimensionalities"). Furthermore, even excitons in bulk black phosphorus (BP)[32] and $TiS_3$ crystal[33] ideally correlate with other vdW materials, as seen in Figure 5. As a result, the comparison in Figure 5 reveals that excitons in vdW materials show 2D-like behavior similar to their monolayer counterparts.

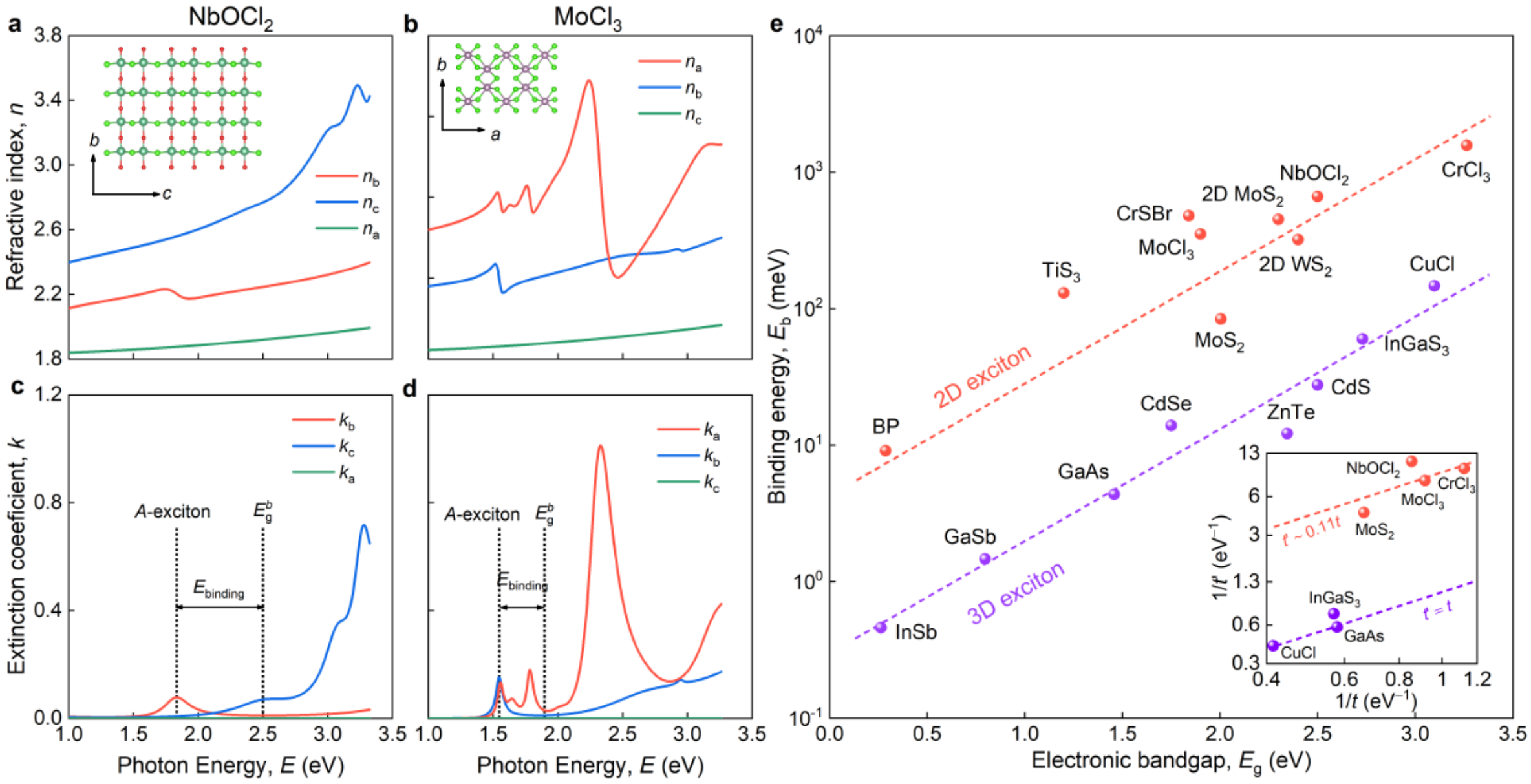

**Figure 4. Layered crystals with giant exciton binding energies.** The refractive index of **a,** $NbOCl_2$ and **b,** $MoCl_3$. The inset shows the crystal structure of these materials. Extinction coefficient of **c,** $NbOCl_2$, and **d,** $MoCl_3$. **e,** The graph could be divided into two regions: "2D exciton" and "3D exciton". The inset shows various compounds in the inverse intraplane ($1/t$)-inverse interplane ($1/t'$) hopping parameters coordinates.

## Discussion

In summary, we positively resolved the long-standing problem of monolayer-like excitonic behavior in layered materials in the example of $CrCl_3$, $NbOCl_2$, and $MoCl_3$. These findings generalize the results for other vdW crystals obtained in earlier works[12,32,33], including BP, $TiS_3$, and $MoS_2$. Moreover, excitons in $CrCl_3$ result from short-order magnetic order in the crystal, offering a potential magnetic control knob of excitonic properties. Together with a facile on-chip integration of vdW crystals[34], it provides an unprecedented platform for exciton-integrated circuits[35] with numerous applications, including polaritonics[36], strong coupling regimes[37], optoelectronics[38], and biosensing[39].

Finally, our findings establish the magnetic origin of excitons in non-magnetic $CrCl_3$ at room temperature. This greatly expands the rich diversity of magnetic excitons, observed in CrSBr[40–42] and $NiPS_3$[43]. Therefore, we anticipate corresponding phenomena for $CrCl_3$, including magnetically confined surface excitons[40], magnetically controlled transport[41], hyperbolic exciton–polariton[42], and manipulation of Coulomb interaction using magnetic order[44]. As a result, our work unveils the unique material platform to create and control nanoscale quantum phenomena in vdW materials.

## Author Contributions

G.E. and T.M. contributed equally to this work. G.E., G.Tselikov, A.V., A.A., S.Z., A.K., K.S.N., and V.S.V. suggested and directed the project. G.A., A. Minnekhanov, G.Tselikov, A.S., M.T., M.K., N.P., A. Mazitov, A. Melentev, E.Z., V.S., I.K., G. Tikhonowski, and D.G. performed the measurements and analyzed the data. T.M., A.M., I.K., and A.K. provided theoretical support. G.E., T.M., A. Mazitov, and A.K. wrote the original manuscript. All authors reviewed and edited the paper. All authors contributed to the discussions and commented on the paper.

## Competing Interests

The authors declare no competing financial interest.

## Acknowledgements

The authors thank Dr. Alexey Tsapenko for the discussions of the article's findings. The work of A. K, T. M. (performing DMFT calculations) was supported by RSF project No 24-12-00186.

## Methods

**Theory.** Theoretical results on the optical properties of $CrCl_3$ were obtained within the BSE@GW framework using VASP code[45]. We used the experimentally characterized crystal structure with the lattice parameters of $a$ = 0.59588 nm, b = 1.0206 nm, $c$ = 0.61138 nm, α = 90°, $\beta$ = 108.495°, $\gamma$ = 90° from reference[23]. The exchange-correlation effects of the electrons in the DFT run were described within the Perdew–Burke–Ernzerhof functional. The behavior of the core electrons was modeled with the projector-augmented wave GW-type pseudopotentials[46]. The first Brillouin zone was sampled with the Γ-centered 6×4×6 mesh. The cutoff energy of the plane waves basis set was 400 eV. We considered three different variations of the Cr spins ordering: ferromagnetic (FM), with the out-of-plane (OOP FM) and the in-plane (IP FM) direction, and so-called A-type antiferromagnetic (AFM) ordering[18] with interlayer AFM spin order and intralayer in-plane FM order. All theoretical calculations started with the non-collinear density functional theory (DFT) run with spin-orbit interaction taken into account to obtain the initial Kohn–Sham orbitals. The convergence threshold of the DFT self-consistent cycle was $10^{-8}$ eV. Next, we estimated the electronic many-body correlation effects and computed the screened Coulomb interaction kernels within the GW approximation[47], and calculated the optical properties without electron-hole interaction. Finally, we calculated the excitonic effects and obtained the excitation energies by solving the Bethe–Salpeter equation (BSE)[48]. The convergence of the excitonic peaks was achieved on taking 20 occupied and 14 virtual bands in the BSE run. Due to the limitations in the treatment of vertex corrections within the self-energy calculations, while calculating the optical absorbance we performed a rigid blue-shift of the absorbance curve by 0.5 eV, following the reasoning of reference[16].

For DFT+DMFT calculations we use Standard solid-state pseudopotentials (SSSP PBEsol Efficiency v1.3.067) without considering spin-orbit interaction. The convergence threshold of the DFT self-consistent cycle was $10^{-8}$. The first Brillouin zone was sampled with the Γ-centered 20×20×20 $k$-points mesh. The resulting band structures were downfolded to Cr d-orbital states and Cl p-orbital states with Wannier90 package[49]. This also allowed us to estimate the hopping parameter $t$ (within the layer), which was taken as the sum of the orbital-averaged contributions from the nearest atoms in the primitive cell within a layer normalized by the number of considered bonds, and the interlayer hopping $t'$, which was obtained as the average contribution of hopping between atoms from different layers (considering all possible combinations within primitive cell in each layer, normalized in the same way as $t$). Next, we estimated the electronic many-body correlation effects within the DMFT approach, realized in the iQIST package[50]. Nonlocal calculation of susceptibilities was performed by solution of the Bethe-Salpeter equations[26] with an account of 60–80 fermionic Matsubara frequencies, including corrections to finite frequency box[51,52].

**Raman and Atomic Force Microscopy Analysis.** The angle-resolved polarized Raman (ARPR) analysis was conducted with an alpha300 RA confocal Raman-Atomic Force Microscope (WITec, Ulm, Germany). A 532 nm laser was employed with a minimal power of 0.05 mW (corresponding to a laser density of ~0.7 mW/μm$^2$) to prevent sample damage. The ARPR spectra were recorded in 10-degree increments (36 points total) in a parallel configuration (with the polarizer aligned parallel to the analyzer). A grating of 600 lines/mm was used, and the backscattered light was collected on a back-illuminated deep depletion CCD detector cooled to −60°C, achieving a spectral resolution of ~1 $cm^{-1}$. Each acquisition had a time of 30 s, repeated 10 times at each angle. Spectra were collected using a 100× objective (Zeiss EC Epiplan-Neofluar, NA 0.9 DIC).

Angle-resolved polarized PL spectra were acquired with the same setup by recording the signal every 5 degrees (72 points) of laser polarizer rotation in a parallel configuration. Each spectrum was collected once with a 1 s acquisition time. Two laser sources were utilized: a 532 nm laser at 0.05 mW, yielding a laser density of ~0.7 mW/$\mu m^2$, and a 785 nm laser set to 1 mW, corresponding to a laser density of ~6 mW/$\mu m^2$. For both wavelengths, a 300 g/mm grating with a broad spectral window was used. Spectra were collected using a 100× objective (Zeiss EC Epiplan-Neofluar, NA 0.9 DIC).

A Linkam T96-S heating stage was employed for temperature-dependent PL measurements. In these experiments, a 532 nm laser with a power of 0.15 mW was used in conjunction with a 50× objective (Zeiss LD EC Epiplan-Neofluar Dic 50× / 0.55), resulting in a laser density of approximately 0.8 mW/$\mu m^2$. A grating of 300 g/mm was utilized, and each spectrum was collected with an integration time of 1 s, averaged over 5 accumulations. To ensure consistent measurements, the PL intensity at each temperature step was normalized to the intensity of the silicon Raman peak, mitigating the effects of focus shifts. The substrate temperature was monitored using a FLIR C5 thermal camera.

Spectral processing, including cosmic ray removal, background subtraction, averaging, and smoothing, was carried out using WITec Project SIX v. 6.2 software.

Atomic Force Microscopy images were recorded in a tapping mode using the WITec alpha300 RA microscope equipped with a NanoWorld ARROW-FMR probe (75 kHz, 2.8 N/m, tip radius < 10 nm).

**Spectroscopic ellipsometry.** Ellipsometry of $CrCl_3$, $NbOCl_3$ and $MoCl_3$ were performed via Accurion EP4 imaging ellipsometer in rotating compensator mode. The samples of $MoCl_3$ and $NbOCl_2$ were aligned along their crystallographic axes to eliminate intermixing of s- and p-polarizations. The measurements were done in a broad spectral range of 250–1700 nm.

**Reflection and transmission microspectroscopy.** Polarized microspectroscopy measurements in the visible range were performed using optical microscopes equipped with a halogen light source polarizer and analyzer. The microscope Zeiss Axio Lab.A1 with the objective “N-Achroplan” 50×/0.8 Pol M27 was utilized for transmittance measurements. For reflectance measurements, the metallurgical microscope RX50M with the objective SOPTOP MPlanFL, ×50/0.8 was used. Transmitted and reflected light were collected from an area less than <20 µm in diameter and directed to the grating spectrometer (Optosky ATP5020P) via an optical fiber (Thorlabs M92L02). A Bruker Vertex 80v Fourier transform infrared spectrometer equipped with a Hyperion 2000 microscope was employed for polarization-resolved infrared spectroscopy. Normal incidence reflection measurements were performed using a standard 15× reflective objective (NA = 0.4). The near-infrared (NIR, 900–1400 nm) setup included a halogen lamp as the light source, a $CaF_2$ beamsplitter, and a mercury cadmium telluride detector.

## Data Availability

The datasets generated during and/or analyzed during the current study are available from the corresponding author upon reasonable request.